\documentclass[prl,10pt,aps,nofootinbib,twocolumn,floatfix,amssymb,amsmath]{revtex4-1}
\usepackage{slashed}
\usepackage{epsfig}

\newcommand{\beqn}{\begin{eqnarray}}
\newcommand{\eeqn}{\end{eqnarray}}
\newcommand{\eq}[1]{(\ref{#1})}

\newcommand{\cL}{{\cal L}}
\newcommand{\cZ}{{\cal Z}}

\newcommand{\cD}{{\cal D}}

\newcommand{\dd}{d}

\newcommand{\cl}{{\mathrm{cl}}}

\newcommand{\Dirac}{\slashed D}

\newcommand{\bs}{\boldsymbol}

\newcommand{\avr}[1]{{\left\langle #1 \right\rangle}}

\begin{document}

\title{Anomalous Transport Due to the Conformal Anomaly}

\author{M.~N.~Chernodub}
\email{maxim.chernodub@lmpt.univ-tours.fr}
\affiliation{CNRS, Laboratoire de Math\'ematiques et Physique Th\'eorique UMR 7350, Universit\'e de Tours, 37200 France}
\affiliation{Soft Matter Physics Laboratory, Far Eastern Federal University, Sukhanova 8, Vladivostok, 690950, Russia}
\affiliation{Department of Physics and Astronomy, University of Gent, Krijgslaan 281, S9, B-9000 Gent, Belgium}

\begin{abstract}
We show that the scale (conformal) anomaly in field theories leads to new anomalous transport effects that emerge in an external electromagnetic field in an inhomogeneous gravitational background. In inflating geometry the QED scale anomaly locally generates an electric current that flows in opposite direction with respect to background electric field (the scale electric effect). In a static spatially inhomogeneous gravitational background the dissipationless electric current flows transversely both to the magnetic field axis and to the gradient of the inhomogeneity (the scale magnetic effect). The anomalous currents are proportional to the beta function of the theory.
\end{abstract}

\date{September 30, 2016}

\maketitle

Anomalous transport phenomena emerge in systems with quantum anomalies that break certain classical symmetries and lead to nonconservation of associated (otherwise classically conserved) currents~\cite{Basar:2012gm,Landsteiner:2012kd}. For example, the axial symmetry of chiral (massless) fermions is broken by the axial anomaly that naturally leads to nonconservation of the axial current at the quantum level~\cite{Shifman:1988zk}:
\beqn
\partial_\mu j^\mu_A = \frac{e^2}{16 \pi^2} F_{\mu\nu} {\widetilde F}^{\mu\nu}\,,
\label{eq:axial:anomaly:j}
\eeqn
where $F_{\mu\nu} = \partial_\mu A_\nu - \partial_\nu A_\mu$ is the field-strength tensor of an Abelian gauge field $A_\mu$ and ${\widetilde F}^{\mu\nu} = \frac{1}{2}  \varepsilon^{\mu\nu\alpha\beta} F_{\alpha\beta}$. 

The simplest anomalous transport laws induced by the axial anomaly~\eq{eq:axial:anomaly:j} are the chiral separation effect (CSE)~\cite{Son:2004tq,Metlitski:2005pr} and the chiral magnetic effect (CME)~\cite{Fukushima:2008xe,Vilenkin:1980fu}:
\beqn
{\bs j}_A = \frac{\mu_V}{2 \pi^2} e {\bs B}\,, \qquad
{\bs j}_V = \frac{\mu_A}{2 \pi^2} e {\bs B}\,,
\label{eq:CSE:j}
\label{eq:CME:j}
\eeqn
that generate, respectively, the axial current ${\bs j}_A$ and the vector current  ${\bs j}_V$ along the axis of the external magnetic field ${\bs B}$ in dense ($\mu_V \neq 0$) and in chirally imbalanced ($\mu_A \neq 0$) medium. The chemical potential $\mu_V$ and the chiral chemical potential $\mu_A$ are thermodynamically conjugated to the total charge density $j^0_V$ and to the axial charge density $j^0_A$, respectively. 

The axial anomaly~\eq{eq:axial:anomaly:j} is also responsible for the density-dependent contributions to the chiral vortical effects~\cite{Vilenkin:1979ui,Son:2009tf} which generate vector and axial currents,
\beqn
{\bs j}_V = \frac{\mu_V \mu_A}{\pi^2} {\bs \Omega} \,, 
\qquad 
{\bs j}_A =\left( \frac{T^2}{6} + \frac{\mu_V^2 + \mu_A^2}{2 \pi^2} \right) {\bs \Omega}\,,
\label{eq:CVE:jVA}
\eeqn
in chiral fluids that rotate with the angular velocity~${\bs \Omega}$. The temperature-dependent $T^2$ part of the rotation-induced axial current in Eq.~\eq{eq:CVE:jVA} is a result of the mixed axial-gravitational anomaly~\cite{Landsteiner:2011cp}:
\beqn
\partial_\mu j^\mu_A = - \frac{1}{384 \pi^2} R_{\mu\nu\alpha\beta} {\widetilde R}^{\mu\nu\alpha\beta}\,,
\label{eq:mixed:anomaly:j}
\eeqn
where ${\widetilde R}^{\mu\nu\alpha\beta}$ is the Riemann curvature tensor of a curved space background and 
${\widetilde R}^{\mu\nu\alpha\beta} = \frac{1}{2} \varepsilon^{\mu\nu\gamma\lambda}  R_{\gamma\lambda}^{\phantom{\gamma\lambda}\alpha\beta}$. Despite that anomaly~\eq{eq:mixed:anomaly:j} is formulated in a curved background, the associated anomalous transport is realized in a flat space too. In the presence of electromagnetic field in a curved background the total divergence of the axial current is given by the sum of the right-hand sides of Eqs.~\eq{eq:axial:anomaly:j} and \eq{eq:mixed:anomaly:j}.

The anomalous transport laws~\eq{eq:CME:j} and \eq{eq:CVE:jVA} are invariant under time inversion $T{:}\ t\to-t$. Since the entropy cannot decrease with time, the $T$ invariance implies that these anomalous currents correspond to reversible processes which do not generate entropy. In other words the anomalous transport laws are nondissipative phenomena. They play an increasingly important role both in condensed matter physics and in high energy physics~\cite{Volovik2003,Kharzeev:2015kna}.

Besides the axial and mixed axial-gravitational anomalies, a class of physically interesting quantum field theories is also subjected to a scale anomaly\footnote{Also known as the dilatation, trace, conformal or Weyl anomaly.} which breaks classical scale invariance of the theory at the quantum level~\cite{Shifman:1988zk}. Since it seems now quite natural to think that at least some quantum anomalies may be associated with certain anomalous transport laws~\cite{Basar:2012gm,Landsteiner:2012kd}, we would like to ask a natural question: does the conformal anomaly lead to a new anomalous transport law?

In this Letter we consider a simplest case of a U(1) gauge theory with one massless Dirac fermion field $\psi$ described by the following Lagrangian:
\beqn
\cL = - \frac{1}{4} F_{\mu\nu} F^{\mu\nu} + {\bar \psi} i \Dirac \psi\,,
\label{eq:L:QED}
\eeqn
where $\Dirac = \gamma^\mu D_\mu$ and $D_\mu = \partial_\mu + i e A_\mu$ is the covariant derivative. This theory does not involve any characteristic length or energy scale since its Lagrangian~\eq{eq:L:QED} possesses only dimensionless coupling $e$. Therefore at a classical level the massless electrodynamics~\eq{eq:L:QED} is invariant under redefinition of the absolute length or energy scales. The corresponding scale transformations are generated by the dilatation current:
\beqn
j_D^\mu = T^{\mu\nu} x_\nu\,,
\label{eq:j:D}
\eeqn
where the (symmetric) energy-momentum tensor
\beqn
T^{\mu\nu} & = & - F^{\mu\alpha} F^\nu_\alpha + \frac{1}{4} \eta^{\mu\nu} F_{\alpha\beta} F^{\alpha\beta} 
\label{eq:Tmunu:QED}\\
& & + \frac{i}{2} {\bar \psi} \left(\gamma^\mu D^\nu + \gamma^\nu D^\mu \right) \psi - \eta^{\mu\nu} {\bar \psi} i \Dirac \psi\,,
\nonumber
\eeqn 
can be obtained by the variation of the action $S$ with respect to the background metric $g_{\mu\nu}$:
\beqn
T^{\mu\nu} (x) = 2 \frac{\delta S}{\delta g_{\mu\nu}(x)} \,,
\qquad
S = \int d^4 x \, \sqrt{-g} \, \cL\,,
\label{eq:Tmunu:dS}
\eeqn
with $g = \det (g_{\mu\nu})$. We restore the flat space-time metric $g^{\mu\nu}{\to}\, \eta^{\mu\nu} = {\mathrm{diag}}\, (+1,-1,-1,-1)$ after the variation.

Classically, the dilatation current~\eq{eq:j:D} has zero divergence because the classical equations of motion imply
\beqn
\partial_\mu j_D^\mu = T^{\alpha}_\alpha\,,
\label{eq:dJ:D}
\eeqn
while the trace of the energy-momentum tensor~\eq{eq:Tmunu:QED} vanishes at the classical level, $T^\alpha_\alpha = 0$. Therefore the classical theory is invariant under the scale transformations. 

However, the scale invariance is broken by quantum fluctuations. Consequently, the quantum expectation value of the right-hand side of Eq.~\eq{eq:dJ:D} is nonzero and, consequently, on a quantum level the dilatation current~\eq{eq:j:D} is no more conserved.

Consider a Weyl scale transformation of the flat metric, $\eta_{\mu\nu} \to g_{\mu\nu}(x)$, with 
\beqn
g_{\mu\nu}(x) = e^{2 \tau(x)} \eta_{\mu\nu}\,,
\label{eq:g:tau}
\eeqn
For small scale factor $\tau(x)$ with $|\tau(x)| \ll 1$ the metric perturbation is $\delta g_{\mu\nu}(x) = 2 \tau(x) \, \eta_{\mu\nu}$ 
and Eq.~\eq{eq:Tmunu:dS} implies:
\beqn
S \to S_{\tau} = S + \int d^4 x \, \tau(x) \, T^\alpha_\alpha(x) + O(\tau^2)\,,
\quad
\label{eq:action:variation}
\eeqn
where $S_\tau$ is the action~\eq{eq:Tmunu:dS} of the theory~\eq{eq:L:QED} in a background of the rescaled flat metrics~\eq{eq:g:tau}. Therefore the expectation value of the trace of the energy-momentum tensor is given by the functional derivative
\beqn
\avr{T^\alpha_\alpha(x)} = \frac{1}{i} \frac{1}{\cZ[A^\cl,\tau]} \frac{\delta  \cZ[A^\cl,\tau]}{\delta \tau(x)}
\eeqn
of the generating functional
\beqn
\cZ[A_\cl,\tau] = \int \! \cD A\,  \cD {\bar \psi} \, \cD \psi \, e^{i S_\tau[A+A^\cl;{\bar \psi}, \psi]}\,,
\label{eq:Z:Acl}
\eeqn
where we have also coupled our system to a background of the classical electromagnetic field~$A_\mu^\cl = A_\mu^\cl(x)$. The latter allows us to express the electric (vector) current of the fermions,
\beqn
j^\mu(x) \equiv j^\mu_V(x) = e {\bar \psi}(x) \gamma^\mu \psi(x)\,,
\label{eq:J:mu}
\eeqn
in terms of a functional derivative
\beqn
\avr{j^\mu(x)} = i \frac{1}{\cZ[A^\cl,\tau]} \frac{\delta  \cZ[A^\cl,\tau]}{\delta A_\mu^\cl(x)}\,.
\label{eq:j:diff}
\eeqn

The scale invariance should generally be broken at the quantum level because the (dimensionless) gauge coupling $e = e(\mu)$ is a function of the (dimensionful) renormalization scale~$\mu$. The Weyl scale transformation~\eq{eq:g:tau} changes the renormalization scale, $\mu \to \mu + \delta \mu$ with $\delta \mu = \mu \delta \tau$ and, consequently, shifts the coupling $e \to e + \delta e$ by $\delta e = \beta(e) \delta \tau$ where
\beqn
\beta(g) = \frac{{\mathrm d}}{{\mathrm d} \ln \mu} \frac{e^2(\mu)}{4 \pi},
\label{eq:beta:g}
\eeqn
is the $\beta$ function of the theory. Therefore at the quantum level the trace of the energy-momentum tensor is nonzero\footnote{
In contrast to the axial anomaly~\eq{eq:axial:anomaly:j} the scale anomaly~\eq{eq:Tmunu:avr} is not a pure one-loop result. Consequently, multiloop corrections may appear in Eq.~\eq{eq:Tmunu:avr}, see Ref.~\cite{Shifman:1988zk} for detailed discussion.},
\beqn
\avr{T^\alpha_\alpha(x)} = \frac{\beta(e)}{2 e} F_{\mu\nu}(x) F^{\mu\nu}(x)\,,
\label{eq:Tmunu:avr}
\eeqn
and the dilatation current~\eq{eq:j:D} is no more conserved  according to Eq.~\eq{eq:dJ:D}. In Eq.~\eq{eq:Tmunu:avr} the field-strength tensor $F_{\mu\nu}$ corresponds to the external background field $A^\cl_\mu$ (for the sake of simplicity we omit hereafter the superscript ``cl'' which refers to the classical background field).

Various aspects of conformal anomalies in hydrodynamics were discussed in Ref.~\cite{ref:conformal}. Below we show that the scale anomaly~\eq{eq:Tmunu:avr} leads to an unexpected (anomalous) contribution to electric current induced by external electromagnetic fields in spatially inhomogeneous or inflating or deflating gravitational backgrounds associated with dilatational perturbations of the form~\eq{eq:g:tau}. In order to demonstrate the essence of the effect we consider the system at zero temperature and zero density so that both usual and chiral chemical potentials are zero, $\mu_V = \mu_A = 0$. In our derivation we follow the logic of Ref.~\cite{Newman:2005as} which we apply to the case of the scale anomaly in the coordinate space. 

The electric current $\avr{j}$ induced by weak external electromagnetic field $A_\mu(x)$ and by small local dilatations of the metric $\tau(x)$ can be expanded in series over these perturbations:
\beqn
\avr{j^\mu} = \avr{j^\mu}_{\mathrm{Kubo}} + \avr{j^\mu}_{\mathrm{dilat}}  + \avr{j^\mu}_{\mathrm{scale}}  + \dots\,. \quad 
\label{eq:J:AD}
\eeqn
The terms in Eq.~\eq{eq:J:AD} are proportional, respectively, to the first powers of $A$ and $\tau$, and to their product $A \tau$. All higher-order terms are denoted by the ellipses.

The first term in Eq.~\eq{eq:J:AD} is given by the standard, linear-response Kubo formula
\beqn
{\left\langle j^\mu(x) \right\rangle}_{\mathrm{Kubo}} = - i \int \dd^4 y \, \Pi^{\mu\nu}(x,y) A_\nu(y)\,,
\label{eq:Kubo:A}
\eeqn
where 
\beqn
\Pi^{\mu\nu}(x,y)  = \avr{j^\mu(x) j^\nu(y)}_0\,,
\label{eq:Pi}
\eeqn
is a two-point correlation function of electric currents. The subscript $0$ in $\avr{\dots}_0$ indicates that the expectation value~\eq{eq:Pi} is calculated in a flat Minkowski space-time in the absence of external perturbations ($A_\mu = 0$, $\delta g_{\mu\nu} = 0$). 

The second term in Eq.~\eq{eq:J:AD} corresponds to a linear response of the current to the pure dilatation~\eq{eq:action:variation},
\beqn
\avr{j^\mu(x)}_{\mathrm{dilat}}
= i \int d^4 y \, \Pi^{\mu}_D(x,y) \tau(y)\,,
\label{eq:Kubo:D}
\eeqn
where 
\beqn
\Pi^{\mu}_D(x,y)  = \avr{j^\mu(x) T^\alpha_\alpha(y)}_0\,,
\label{eq:Pi:jD}
\eeqn
is a two-point correlation function of the electric current~\eq{eq:J:mu} and the trace of the energy-momentum tensor~\eq{eq:Tmunu:QED}. The correlation function~\eq{eq:Pi:jD} can be calculated by varying the anomalous expectation value~\eq{eq:Tmunu:avr} with respect to external electric field $A_\mu$ in a manner of Eq.~\eq{eq:j:diff}, and setting $A_\mu = 0$ after the variation. Since the anomaly~\eq{eq:Tmunu:avr} is quadratic in gauge field $A_\mu$ the correlation function~\eq{eq:Pi:jD} is zero. Therefore the electric current~\eq{eq:Kubo:D}, induced by the dilatation, is vanishing in the linear response approximation, $\avr{j^\mu(x)}_{\mathrm{dilat}} \equiv 0$.

In our Letter we are mainly interested in the third term in Eq.~\eq{eq:J:AD}. This term describes a scale-anomalous contribution to the expectation value of the electric current. It corresponds to a mixed gauge-gravitational response in the double-linear approximation that includes one power of the electromagnetic potential $A_\mu$ and one power of the scale factor~$\tau$. According to Eq.~\eq{eq:Kubo:A}
\beqn
\avr{j^\mu(x)}_{\mathrm{scale}} = \int \dd^4 y \int \dd^4 z \, \Pi^{\mu\nu}_D(x,y;z) A_\nu(y) \tau (z),\ \quad
\label{eq:Jmu:A:sigma}
\eeqn
where the three-point function 
\beqn
\Pi^{\mu\nu}_D(x,y;z) = \avr{j^\mu(x) j^\nu(y) T^\alpha_\alpha(z)}_0  \,,
\label{eq:Pi:D}
\eeqn
can be evaluated by applying twice a functional differentiation with respect to the background gauge field $A_\mu$ to the right-hand side of the scale anomaly relation~\eq{eq:Tmunu:avr}:
\beqn
& & \Pi^{\mu\nu}_D(x,y;z) = - \frac{\delta^2 \avr{T^\alpha_\alpha(z)}}{\delta A_\mu(x) \delta A_\nu(y)} {\biggl |}_{{}_{g_{\mu\nu} \to \eta_{\mu\nu}}^{A_\mu \to 0}} 
\label{eq:Pi:D:calculated}
\\
& & \quad  = - \frac{2 \beta(e)}{e} \left(\eta^{\mu\nu} \eta^{\alpha\beta} - \eta^{\mu\beta} \eta^{\nu\alpha}\right) \frac{\partial^2 \delta(x-z) \delta(y-z)}{\partial x^\alpha \partial y^\beta}\,. 
\nonumber 
\eeqn

Substituting Eq.~\eq{eq:Pi:D:calculated} into Eq.~\eq{eq:Jmu:A:sigma} one gets the anomalous electric current 
generated by the scale anomaly~\eq{eq:Tmunu:avr} in the presence of both the scale dilatation $\tau$ of the metric~\eq{eq:g:tau} and the background electromagnetic field~$A^\mu$:
\beqn
\avr{j^\mu(x)}_{\mathrm{scale}} = \frac{2 \beta(e)}{e} \Bigl[- F^{\mu\nu}(x) \partial_\nu \tau(x) + \tau(x) j_\cl^\mu(x)  \Bigr]. \qquad
\label{eq:J:scale}
\eeqn

The first term in Eq.~\eq{eq:J:scale} is proportional to the electromagnetic field $F^{\mu\nu}$, which is induced by the classical electric current $j_\cl^\mu = - \partial_\nu F^{\mu\nu}$. The classical current makes the local contribution to the anomalous electric current given in the second term of Eq.~\eq{eq:J:scale}. The presence of both terms guarantees that the anomalously generated current~\eq{eq:J:scale} is conserved: 
$\partial_\mu \avr{j^\mu(x)}_{\mathrm{scale}} = 0$.

Our result~\eq{eq:J:scale} is obtained via  the three-point function~\eq{eq:Pi:D} which is defined and calculated in the flat Min\-kow\-ski space-time. Thus we do not expect a met\-ric-de\-pen\-dent renormalization of the current~\eq{eq:J:scale} in the adopted linear order in gravitational perturbation. The same property is shared by a contribution to the chiral vortical effect coming from the axial-gravitational anomaly~\cite{Landsteiner:2011cp}, which can also be calculated in the flat space in a linear-response approximation in metric $g_{0i}$.

We are interested in properties of the anomalous electric current far from the classical sources. Therefore, setting the classical current to zero in the region of the dilatation, $j_\cl^\mu = 0$, we get, from Eq.~\eq{eq:J:scale},
\beqn
\avr{j^\mu(x)}_{\mathrm{scale}} = - \frac{2 \beta(e)}{e} F^{\mu\nu}(x) \partial_\nu \tau (x)\,.
\label{eq:J:scale:2}
\eeqn

In components, the anomalous current and the anomalous charge generated by the scale anomaly~\eq{eq:J:scale:2} in the background of the electric field $\bs E$ and the magnetic field $\bs B$ are, respectively, as follows:
\beqn
\avr{{\bs j(x)}}_{\mathrm{scale}} & = & \sigma(x) {\bs E}(x) + {\bs F}(x) \times {\bs B}(x)\,,
\label{eq:Ohm:1:naive}\\
\avr{j^0(x)}_{\mathrm{scale}} & = & {\bs F}(x) \cdot {\bs E}(x) \,, 
\label{eq:Ohm:2:naive}
\eeqn
where 
\beqn
\sigma(t, {\bs x}) & = & - \frac{2 \beta(e)}{e} \frac{\partial \tau(t,{\bs x})}{\partial t}\,,
\label{eq:sigma:anomalous} \\
{\bs F}(t, {\bs x}) & = & \frac{2 \beta(e)}{e} {\bs \nabla} \tau(t, {\bs x})\,,
\label{eq:F:anomalous}
\eeqn
and $\tau(x)$ is the local scale factor of the flat metric~\eq{eq:g:tau}. 

The scalar quantity $\sigma$, given by Eq.~\eq{eq:sigma:anomalous} plays a role of an anomalous Ohm's conductivity. Indeed, in a spatially uniform (${\bs \nabla} \tau \equiv 0$) background $g_{\mu\nu} = e^{2 \tau(t)} \eta_{\mu\nu}$ with a time-dependent scale factor $\tau = \tau(t)$ 
the scale anomaly generates the anomalous electric current~\eq{eq:Ohm:1:naive} which takes precisely the functional form of Ohm's law:
\beqn
\avr{{\bs j}(t, {\bs x})}_{\mathrm{scale}} = \sigma(t) {\bs E}(t, {\bs x}) \qquad {\mathrm{for}} \quad {\bs \nabla} \tau = 0\,.
\label{eq:Ohm}
\eeqn
Equations~\eq{eq:sigma:anomalous} and \eq{eq:Ohm} describe the scale electric effect (SEE): the scale anomaly generates the local electric current in the background of the external electric field in a space-time with a time-dependent scale factor.

\begin{figure}[!thb]
\begin{center}
\includegraphics[scale=0.5,clip=true]{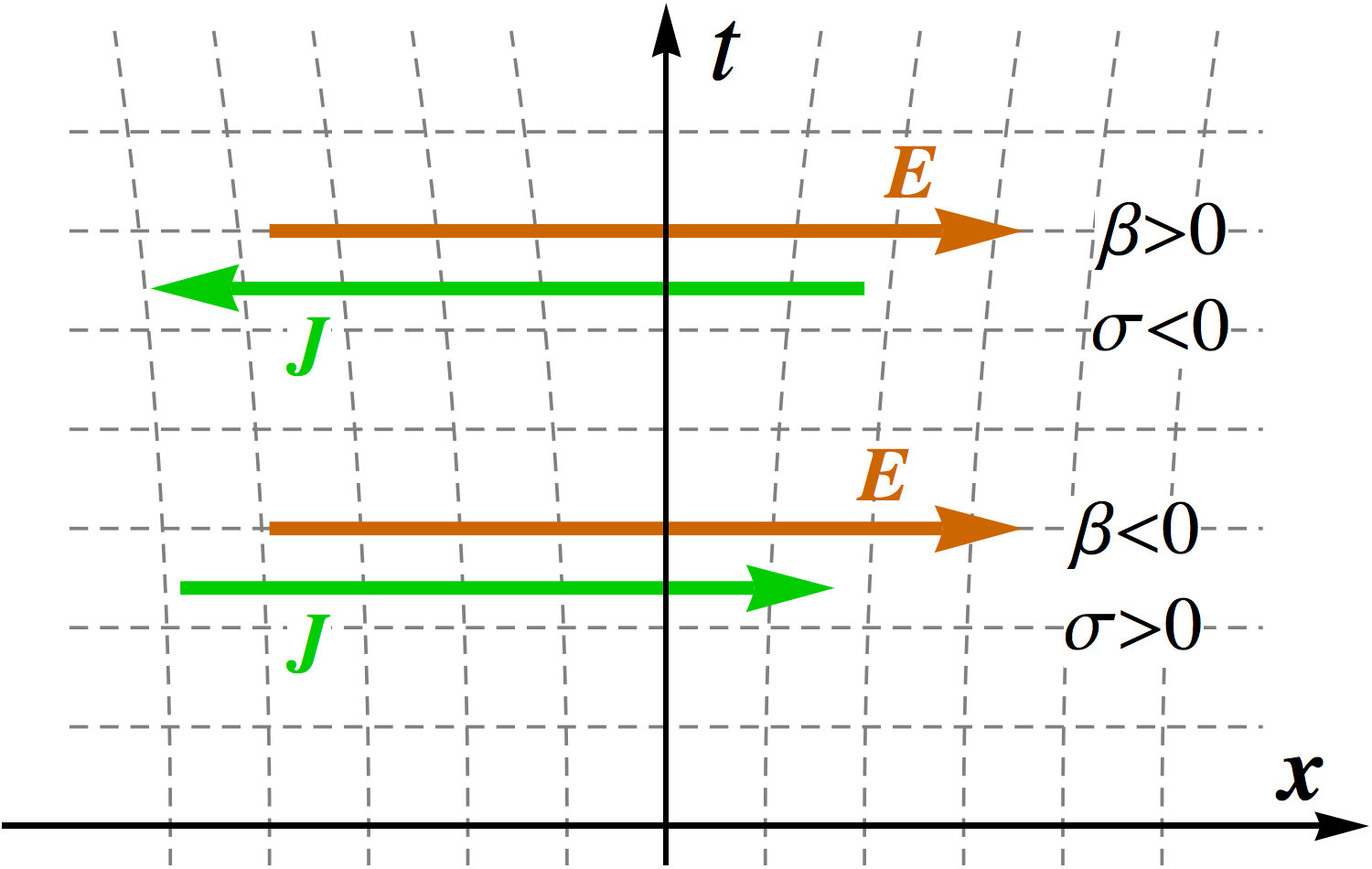}
\end{center}
\vskip -3mm
\caption{The scale electric effect (SEE): in an inflating ($\partial_t \tau {>} 0$) gravitational background \eq{eq:g:tau} the scale (conformal) ano\-ma\-ly~\eq{eq:Tmunu:avr} generates the electric current ${\bs J}$ along the electric field ${\bs E}$ according to Eqs.~\eq{eq:Ohm} and \eq{eq:sigma:anomalous}. The direction of the current depends on the sign of the $\beta$ function.}
\label{fig:Ohm}
\end{figure}

The SEE~\eq{eq:Ohm} emerges in an open, expanding (or contracting) system which has an explicit arrow of time. Consequently, the SEE does not conserve entropy and does not, in general, describe a dissipationless phenomenon contrary to the chiral anomalous transport effects~\eq{eq:CSE:j} and~\eq{eq:CVE:jVA}. The power $P = \avr{{\bs j}}_{\mathrm{scale}} \cdot {\bs E} = \sigma E^2$ dissipated by the anomalous electric current~\eq{eq:Ohm} per unit volume may take both positive and negative values because the anomalous conductivity~\eq{eq:sigma:anomalous} may be both a positive and negative quantity, respectively. As a result, in this {\emph {open}} system the scale electric effect~\eq{eq:Ohm} may not only heat the system but it may also cool it by absorbing heat. We illustrate the SEE~\eq{eq:Ohm} in Fig.~\ref{fig:Ohm}.

\begin{figure}[!thb]
\begin{center}
\includegraphics[scale=1.3,clip=true]{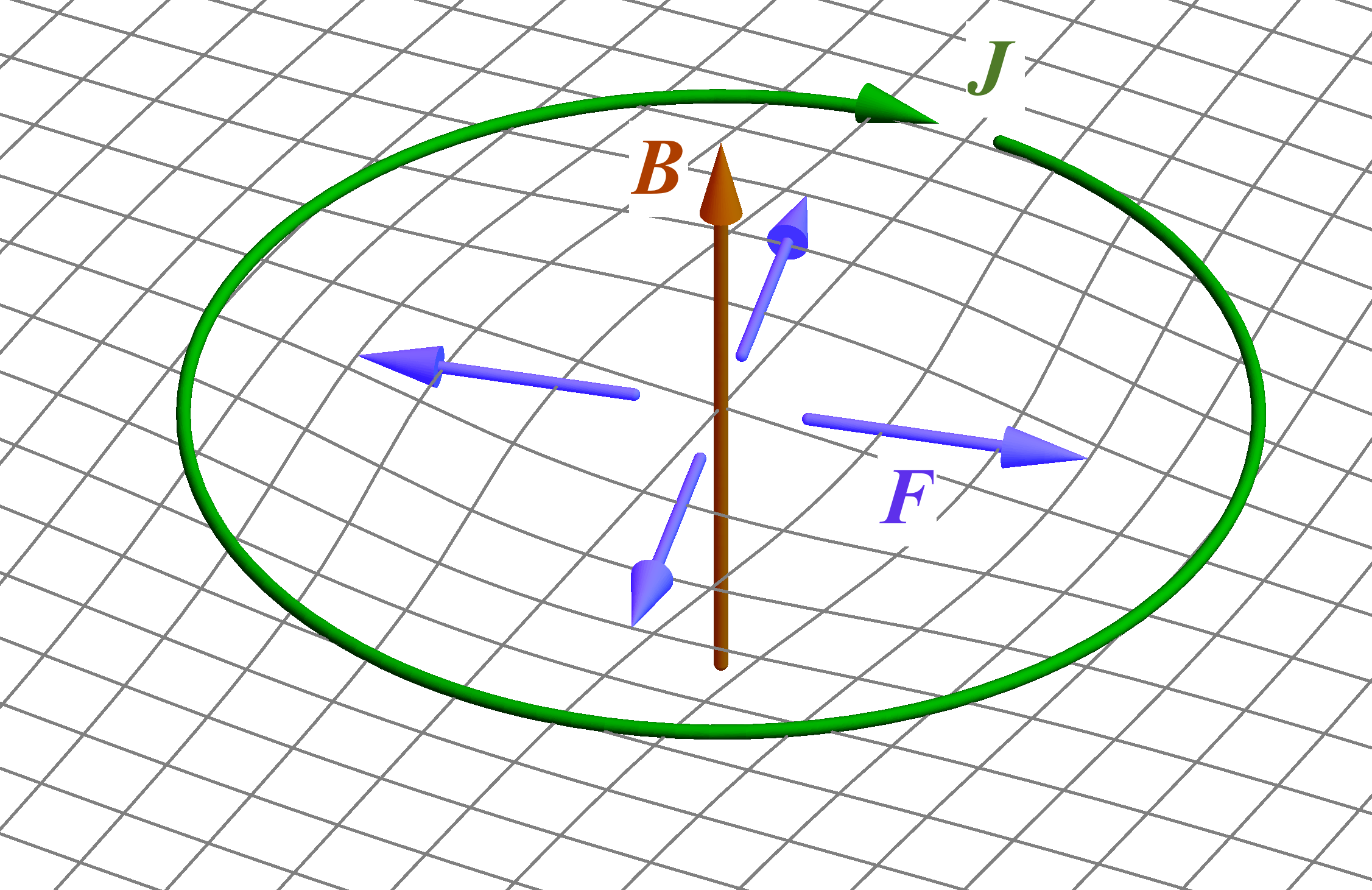} 
\hskip 2mm
\includegraphics[scale=1.45,clip=true]{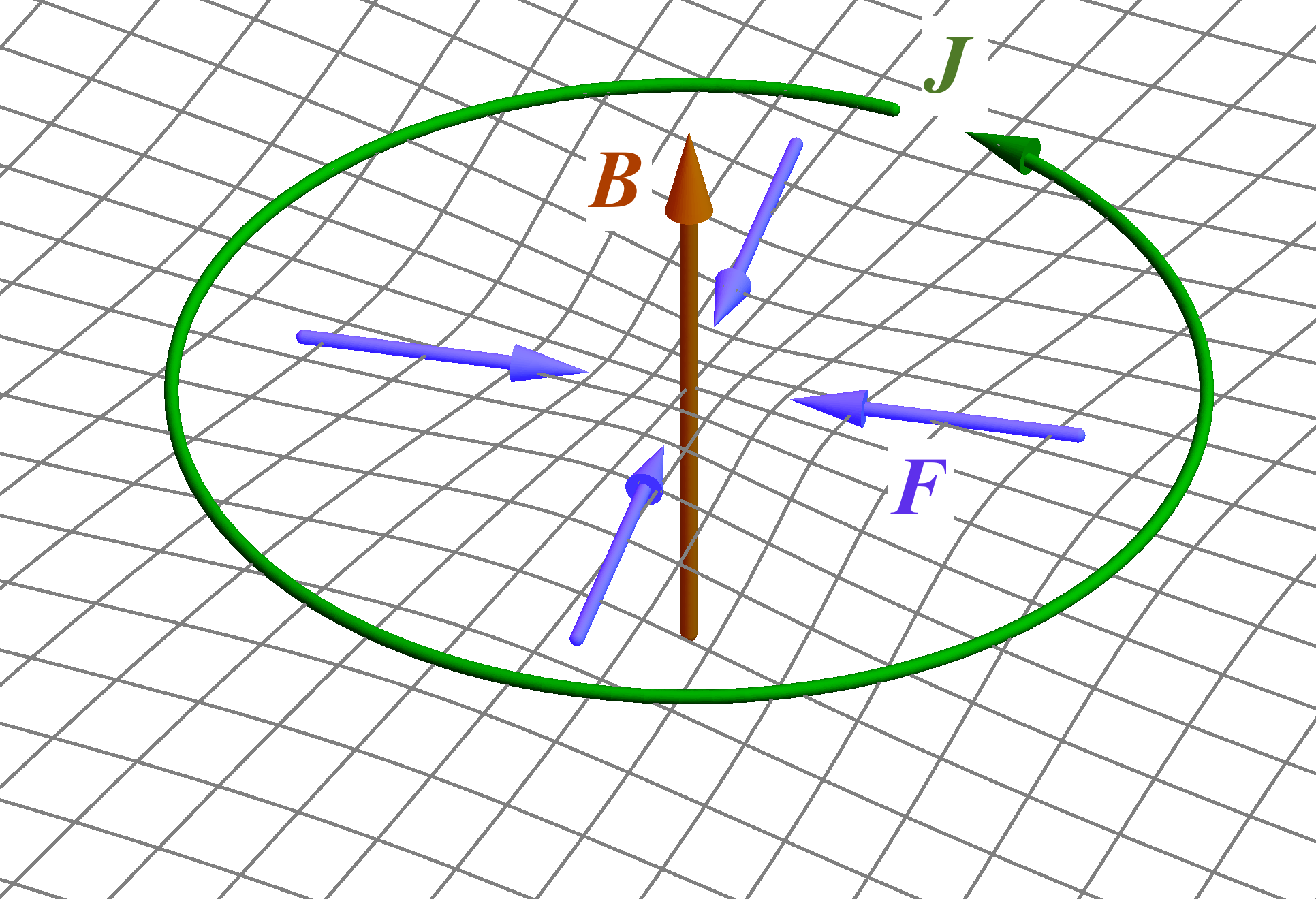}
\end{center}
\vskip -3mm
\caption{The scale magnetic effect (SME): in a spatially nonuniform gravitational background the scale (conformal) ano\-maly~\eq{eq:Tmunu:avr} generates the electric current ${\bs J}$ cir\-cum\-navigating the external magnetic field ${\bs B}$ according to Eq.~\eq{eq:Hall-like}. The gradient of the scale factor ${\bs F}$ is given in Eq.~\eq{eq:F:anomalous}.}
\label{fig:Magnetic}
\end{figure}

The anomalous current~\eq{eq:Ohm:1:naive} has also a contribution coming from the magnetic field ${\bs B}$. This part may only appear due to local spatial inhomogeneities of the scale factor $\tau(x)$, which are encoded in the vector quantity ${\bs F}$ in Eq.~\eq{eq:F:anomalous}. According to Eqs.~\eq{eq:Ohm:1:naive} and \eq{eq:F:anomalous}, in a nonuniformly stretched static space-time the scale anomaly generates the electric current which is transversal both to the direction of magnetic field ${\bs B}$ and to the gradient of the spatial inhomogeneity~${\bs F}$:
\beqn
\avr{{\bs j}(t, {\bs x})}_{\mathrm{scale}} & = & {\bs F}({\bs x}) \times {\bs B}(t, {\bs x}) \quad {\mathrm{for}} \quad \partial_t \tau = 0\,.
\label{eq:Hall-like}
\eeqn
Equations~\eq{eq:F:anomalous} and \eq{eq:Hall-like} describe the scale magnetic effect (SME): the scale anomaly generates the local electric current in background of external magnetic field in a space-time with a spatially dependent scale factor (Fig.~\ref{fig:Magnetic}).

The electric current generated by the scale magnetic effect~\eq{eq:Hall-like} flows without dissipation similarly to the current~\eq{eq:CME:j} generated by the chiral magnetic effect. However, there are major differences between the SME~\eq{eq:Hall-like} and the axial-anomalous transport effects~\eq{eq:CME:j}: (i) the scale anomaly generates the electric current via the SME in the vacuum state while the CME is realized in matter only; (ii) the SME electric current~\eq{eq:Hall-like} is transverse to the direction of magnetic field while in the CME the magnetic field and the current are parallel to each other.

In the presence of external electric field the scale anomaly should also lead to concentration of electric charge~\eq{eq:Ohm:2:naive} at spatial inhomogeneities of the metric~\eq{eq:F:anomalous}. 

Using a one-loop QED $\beta$ function for one species of a Dirac fermion~\eq{eq:L:QED},
$\beta^{{\text{1-loop}}}_{{\text{QED}}}(e) = e^3/(12 \pi^2)$
we get the anomalous transport coefficients~\eq{eq:sigma:anomalous} and \eq{eq:F:anomalous}:
\beqn
\sigma = - \frac{e^2}{6 \pi^2} \frac{\partial\tau}{\partial t}\,, 
\qquad \quad
{\bs F} & = & \frac{e^2}{6 \pi^2}  {\bs \nabla} \tau\,,
\eeqn
where $\tau(x)$ is the local scale factor of the flat metric~\eq{eq:g:tau}.

In an expanding geometry with a generic homogeneous isotropic metric $d s^2 = dt^2 - a^2(t) d {\bs x}^2$ the conductivity of each species of massless charged fermions gets an anomalous contribution (we restore $\hbar$ and $c$): 
\beqn
\sigma_{\text{QED}}(t) = - \frac{e^2 H(t)}{6 \pi^2 \hbar c}\,, 
\label{eq:sigma:Hubble}
\eeqn
where $H(t) = \dot{a}(t)/a(t)$ is the Hubble parameter. The derivation of Eq.~\eq{eq:sigma:Hubble} assumes that the scale factor $\tau$ is small so that the scale factor $a(t)$ is close to unity.

Equation~\eq{eq:sigma:Hubble} implies that the inflating vacuum of massless fermions should have -- due to the scale anomaly~\eq{eq:Tmunu:avr} -- a nonzero {\it negative} conductivity in theories with positive beta functions, $\beta > 0$. This conclusion agrees with the findings of Refs.~\cite{Hayashinaka:2016qqn} and~\cite{Kobayashi:2014zza}, where the electric conductivity induced by, respectively, fermionic and bosonic Schwinger effects, was calculated in inflating (de Sitter) space-time. In particular, in a weak-field limit, $|eE| \ll H^2$, the leading term in the fermionic Schwinger effect~\cite{Hayashinaka:2016qqn} is 4 times bigger than its bosonic counterpart~\cite{Kobayashi:2014zza} in agreement with the relation $\beta^{{\text{1-loop}}}_{{\text{QED}}} = 4 \beta^{{\text{1-loop}}}_{{\text{sQED}}}$ between usual and scalar QED beta functions.\footnote{The field theory situation is still somewhat unclear since fer\-mio\-nic and bosonic conductivities induced by the Schwinger effect~\cite{Kobayashi:2014zza,Hayashinaka:2016qqn} contain explicit contributions from our generic conformal ($m{=}0$) formula~\eq{eq:sigma:anomalous} up to a common logarithmic prefactor $\log(H^2/m^2)$. The existence of this infrared-divergent factor is questioned for the massless case ($m{=}0$) where the adiabatic regularization of~\cite{Hayashinaka:2016qqn,Kobayashi:2014zza} may not be applicable~\cite{Kobayashi:2014zza} (in exactly conformal case no infrared divergences are expected to arise~\cite{Parker:1974qw}).} Thus, in the particular case of ho\-mo\-ge\-neous and isotropic inflation the scale-anomalous conductivity can be associated with the Schwinger pair production. 

The sign of the Lorentz invariant $F_{\mu\nu} F^{\mu\nu} \propto B^2 - E^2$ determines whether the electromagnetic background is magnetically ($|B| > |E|$) 
or electrically ($|E| > |B|$) dominated. In the former (latter) case the pure SME (SEE) is realized in the reference frame in which the electric (magnetic) field vanishes. In a general frame both effects are present, and the induced current is given by Eqs.~\eq{eq:Ohm:1:naive}--\eq{eq:F:anomalous}. The magnetic field dominance is required for the existence of a stable vacuum.

Summarizing, we have shown that in theories with electrically charged massless particles the scale anomaly leads to new transport effects: the scale electric effect (SEE) and the scale magnetic effect (SME) given by Eqs.~\eq{eq:Ohm} and \eq{eq:Hall-like}, respectively. The SEE implies that in inflating geometry the QED scale anomaly generates electric current~\eq{eq:Ohm} flowing in the opposite direction to the electric field, thus exhibiting a negative conductivity, Eq.~\eq{eq:sigma:Hubble}. The SME implies that in a static but spatially inhomogeneous conformal gravitational background the dissipationless electric current flows transversely both to the magnetic field axis and to the gradient of the inhomogeneity, Eq.~\eq{eq:Hall-like}. The generated electric currents are proportional to the appropriate $\beta$ function. One can expect that the scale-anomalous transport effects~\eq{eq:Ohm:1:naive}--\eq{eq:F:anomalous} are quite generic phenomena because the anomalous term~\eq{eq:Tmunu:avr} -- which is our starting point -- is present in wide varieties of physical models involving fermionic and/or bosonic degrees of freedom. They may also presumably be realized in solid state materials possessing relativistic quasiparticles, such as strained graphene~\cite{ref:strained:graphene} or elastically deformed Weyl/Dirac semimetals~\cite{ref:deformed:semimetals}. 

The author is grateful to G.~Gibbons, K.~Landsteiner, M.~Vozmediano and M.~Zubkov for interesting discussions, and to T.~Kalaydzhyan, Takeshi Kobayashi, E.~Megias, M.~Valle and A.~Vilenkin for valuable communications. The author acknowledges hospitality and support of IFT-UAM/CSIC, Madrid, as associate researcher under the Centro de Excelencia Severo Ochoa Programme Grant No. SEV-2012-0249.

\end{document}